\documentstyle[11pt,newpasp,twoside,epsf,natbib,url]{article}
\def\msun{M$_{\sun}$}
\def\A{{\bf A}}
\def\B{{\bf B}}

\def\edcomment#1{\iffalse\marginpar{\raggedright\sl#1\/}\else\relax\fi}
\reversemarginpar

\begin{document}

\title{GBT Exploratory Time Observations of the Double-Pulsar System
  PSR~J0737$-$3039}

\author{S.~Ransom$^1$, P.~Demorest$^2$, V.~Kaspi$^1$,
  R.~Ramachandran$^2$, \& D.~Backer$^2$}

\affil{$^1$Dept. of Physics, McGill University, 3600 University St.,
  Montreal, QC H3A~2T8, Canada}
\affil{$^2$Dept. of Astronomy and Radio Astronomy Laboratory,
  University of California at Berkeley, 601 Campbell Hall 3411,
  Berkeley, CA 94720}

\begin{abstract}
  We report results from Exploratory Time observations of the
  double-pulsar system PSR~J0737$-$3039 using the Green Bank Telescope
  (GBT).  The large gain of the GBT, the diversity of the pulsar
  backends, and the four different frequency bands used have allowed
  us to make interesting measurements of a wide variety of phenomena.
  Here we briefly describe results from high-precision timing,
  polarization, eclipse, scintillation velocity, and single-pulse
  work.
\end{abstract}

\section{Introduction}

In December 2003, we proposed for and were awarded Exploratory Time to
observe the spectacular double-pulsar system PSR~J0737$-$3039
\citep[hereafter 0737;][and contributions to this volume from Burgay,
Manchester, Kramer, and others]{bdp+03,lbk+04} as part of the NRAO
Rapid Response Science program\footnote{\tt
  http://www.vla.nrao.edu/astro/prop/rapid/}.  We observed 0737 five
times at four different frequencies (427, 2$\times$820, 1400, and
2200\,MHz) using the Berkeley-Caltech Pulsar Machine (BCPM) and the
Green Bank Pulsar Processor (GBPP; which measures full polarization
information).  For two of the observations (one each at 427 and 820
MHz) we also used the new GBT Spectrometer SPIGOT card (a
correlator-based instrument that outputs lags at 25\,MB/s) in some of
its first scientific observations.  These data are public and can be
obtained from NRAO.

\section{High-precision Timing}
\label{sec:timing}

\begin{figure}
  {\centering
    \leavevmode
    \epsfxsize=.45
    \textwidth
    \epsfbox{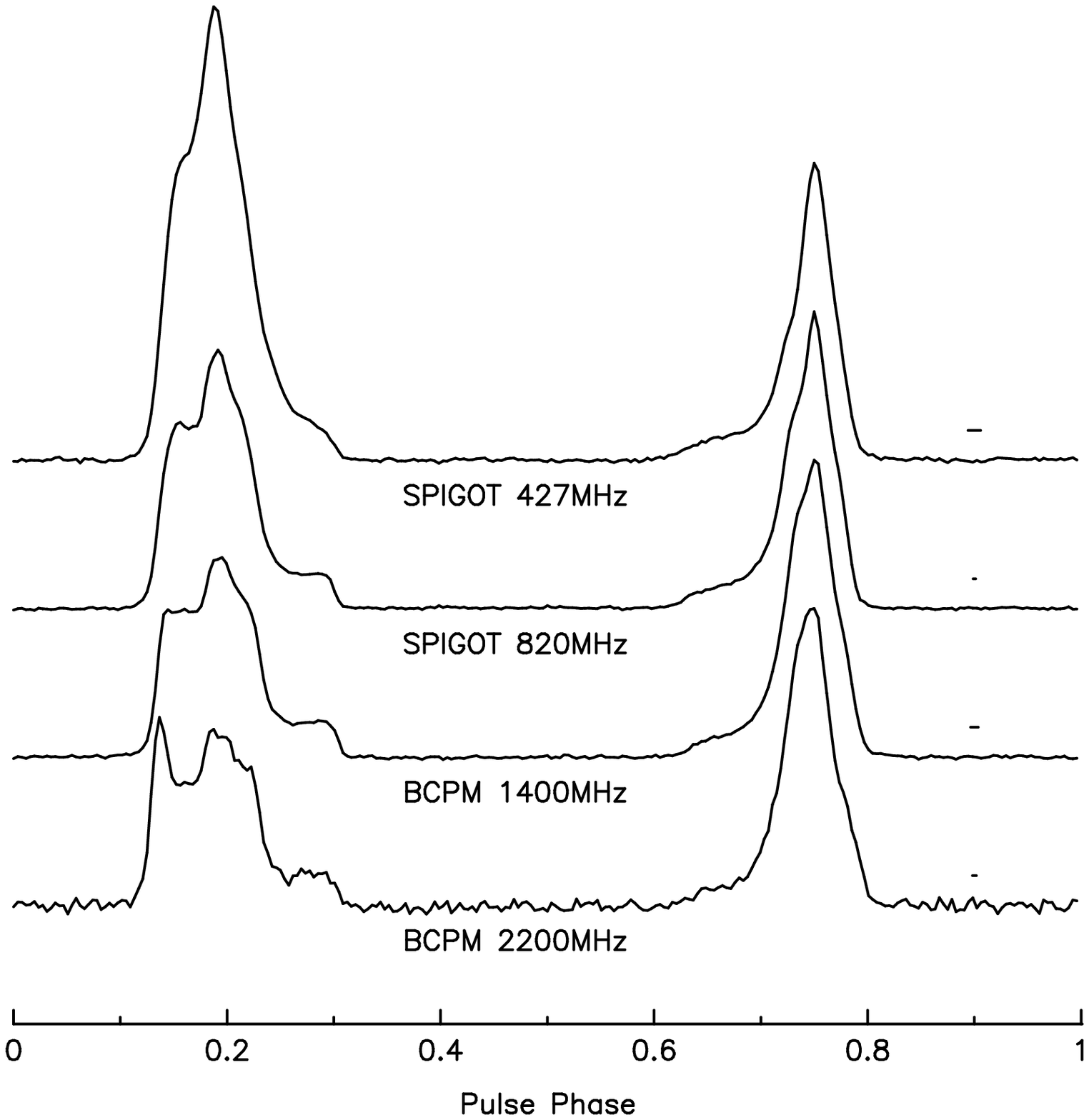}
    \hfil
    \epsfxsize=.45
    \textwidth
    \epsfbox{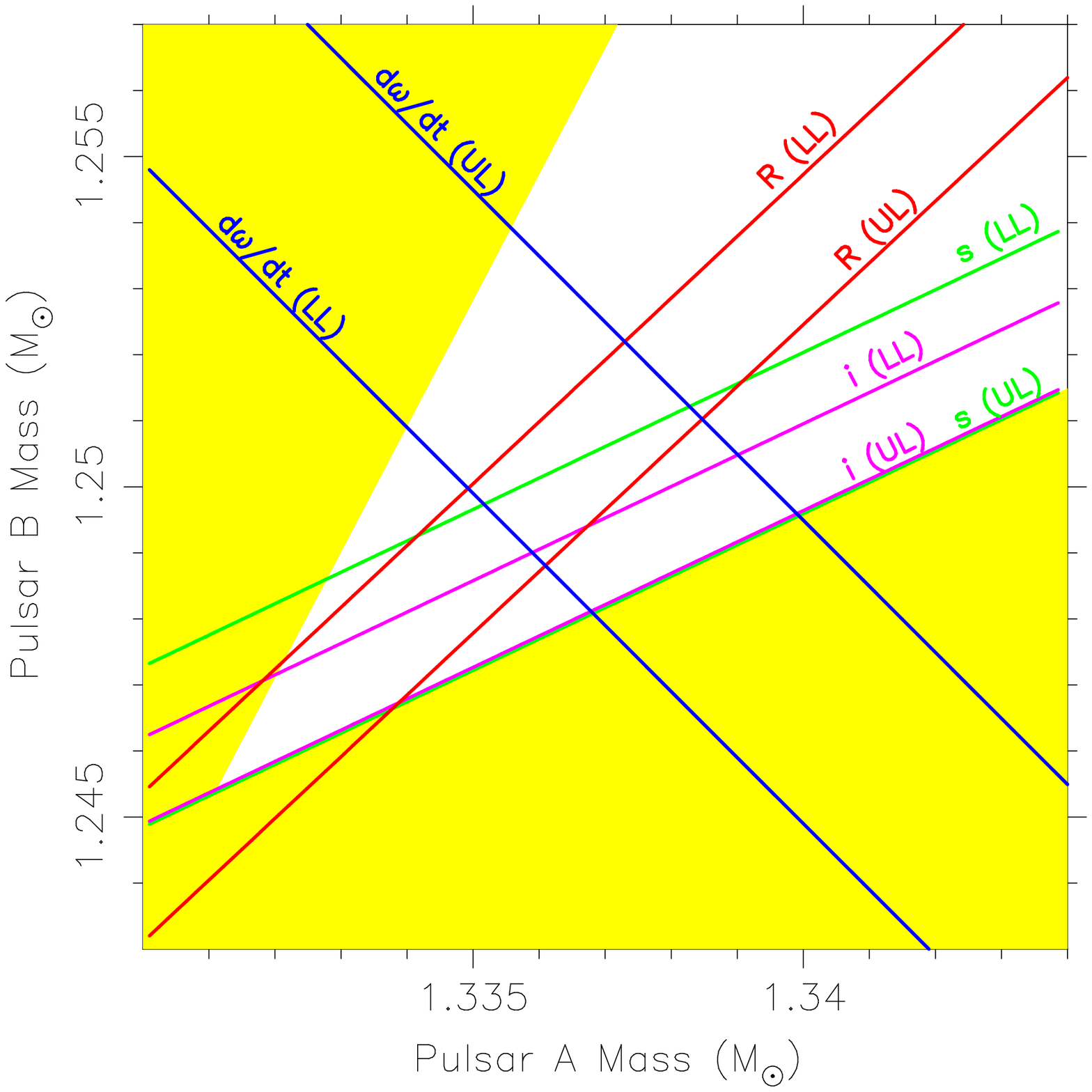}}
  \caption{
    (Left) Average pulse profiles of \A\ from the GBT.  Changes in the
    relative pulse amplitudes with observing frequency are obvious.
    The short lines above and to the right of the profiles depict the
    time resolution of the data. (Right) A diagram of the $M_A$
    vs.~$M_B$ plane \citep[after][]{lbk+04} based on the GBT data
    (except for the advance of periastron $\dot\omega$).  The shaded
    region requires an inclination $i>90\deg$ and is therefore not
    allowed.  The other plotted parameters are the mass ratio $R\equiv
    m_A/m_B$, the Shapiro delay ``shape'' parameter $s\equiv\sin i$,
    and the $i$ as determined using both timing (\S\ref{sec:timing})
    and scintillation measurements (\S\ref{sec:scint}).  Lower and
    upper limits on the various parameters are labeled with LL or UL
    respectively.
    \label{fig:profs}}
\end{figure}

The high sensitivity of the GBT allowed us to measure high
signal-to-noise ratio (SNR) pulse profiles from PSR~J0737$-$3039A
(hereafter \A, see Fig.~\ref{fig:profs}) at each observing frequency,
precise ($\la$11\,$\mu$s) pulse arrival times (TOAs) from \A\ in
1$-$2\,min integrations, and TOAs throughout $\sim$80\% of the orbit
for PSR J0737$-$3039B (hereafter \B).  Without a long timing baseline
we could only measure ``local'' parameters, such as the Shapiro delay
``shape'' $s\equiv\sin i$ for \A, and better projected orbital
semi-major axes $x\equiv a\sin i/c$ for the pulsars.

For \A\, we used the \citet{lbk+04} ephemeris as our base and
extracted TOAs from the high SNR 820- and 1400-MHz BCPM and SPIGOT
observations.  Using {\tt
  TEMPO}\footnote{\url{http://pulsar.princeton.edu/tempo}}, we
determined $s$=0.99962$^{+0.00038}_{-0.00095}$ (corresponding to
$i$=88\fdg4$^{+1.6}_{-1.4}$) and $x_A$=1.4150342(82)\,lt-sec with a
post-fit RMS of $\sim$11\,$\mu$s (the SPIGOT data alone had an RMS of
$\la$7\,$\mu$s).  The errors for $s$ and $x_A$ were estimated using a
bootstrap analysis.  We note that we had to fit out apparent linear
arrival time drifts of order tens of $\mu$s during each of the
observations.  We believe that these (likely) systematic drifts could
result from uncorrected variations in the linearly polarized portion
of the pulse profile due to the rotation of the receiver feed with
respect to the sky and gain differences between the two orthogonally
polarized IFs.  See \citet{rkr+04} for more details of the \A\ timing.

\begin{figure}
  \plotfiddle{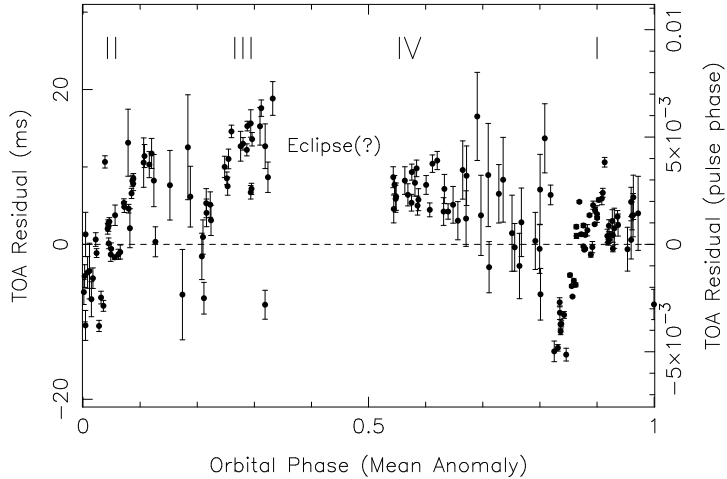}{5.8cm}{0}{50}{50}{-160}{-15}
  \caption{
    Post-fit timing residuals as a function of orbital phase for \B.
    The only fit parameters were the projected semi-major axis $x_B$
    and the spin frequency.  The TOAs are from BCPM data from all five
    GBT observations of 0737.  Systematic drifts in the TOAs are
    obvious, even after using orbital-phase-dependent profile
    templates to extract the TOAs.  The ``bright'' portions of the
    orbit as defined in \citet{drb+04} and \citet{rdr+04} are labeled
    as I, II, III, and IV.  The possible ``eclipse'' of \B\ is also
    indicated (see \S\ref{sec:pulses}).
    \label{fig:Btiming}}
\end{figure}

To time \B, we determined a set of 29 template profiles (all
referenced to the same gaussian component) as a function of the
orbital phase (since \B's pulse profile changes systematically
throughout the orbit) and extracted TOAs from each of the BCPM
observations.  Using the \citet{lbk+04} timing solution as a starting
point, we used {\tt TEMPO} to fit only for $x_B$ and the local spin
frequency.  The post-fit residuals show significant systematics as a
function of orbital phase (see Fig.~\ref{fig:Btiming}), implying that
there may be no stable timing reference point in the varying profile
(i.e. \A's wind likely affects the radio beam of \B\ as a function of
orbital phase).  If we multiply {\tt TEMPO}'s formal errors on $x_B$
by 10.3 (the square-root of the reduced-$\chi^2$ of the fit) to try
and account for some of these systematics, we get
$x_B$=1.5126(14)\,lt-sec and a mass ratio $R\equiv m_A/m_B = x_B/x_A =
1.06893(97)$, which is a factor of $\sim$6 more precise than reported
in \citet{lbk+04}.  A $M_A$ vs. $M_B$ diagram based on the GBT data
but using the reported value of $\dot\omega$ from \citet{lbk+04} is
shown in Fig.~\ref{fig:profs}.

\begin{figure}
  \plotone{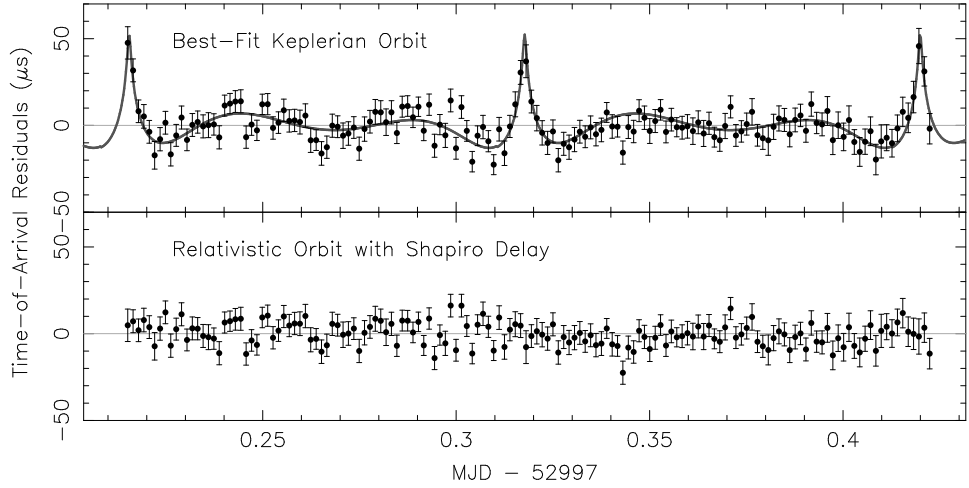}
  \caption{
    Timing residuals from the 820\,MHz SPIGOT observation of \A\ 
    before (top) and after (bottom) the addition of a Shapiro delay
    ``shape'' fit for a 1.25\,\msun\ companion.  The Shapiro delay is
    easily detected during {\em each} orbit. The post-fit RMS
    residuals for these data alone are $\la$7\,$\mu$s.
    \label{fig:shapiro}}
\end{figure}

\section{Polarization}
\label{sec:polar}

The GBPP coherently dedisperses and folds full-Stokes data on a single
pulsar in real time.  The most useful polarimetric data sets were
those taken at 820\,MHz: \B\ on 2003 Dec 19 and \A\ on 2003 Dec 24.
We used the full track on \A\ to perform polarization self calibration
of the data.  This calibration fit included terms accounting for
channel gain difference, feed cross coupling, feed ellipticity, and
cable delays.  The RMS average of all such off-diagonal Mueller matrix
elements was 0.085, so we are confident that our calibration is fairly
good.  We also used measurements of the observatory correlated noise
source and observations of PSR~B1929$+$10 to confirm our calibration.

\begin{figure}
  \plotfiddle{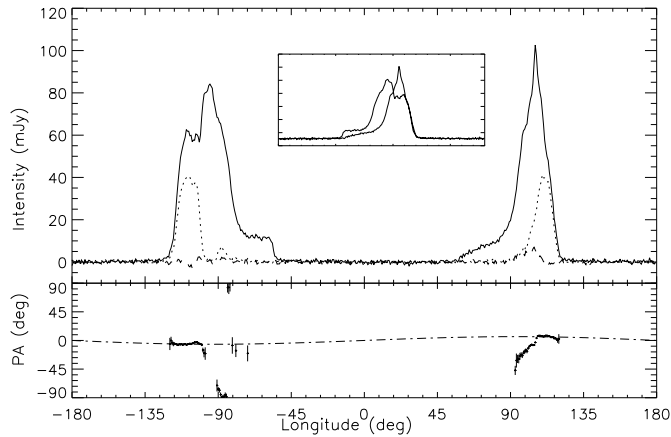}{5.3cm}{0}{60}{60}{-130}{0}
  \caption{
    (Top) GBPP 820\,MHz profiles of \A.  The dotted and dashed lines
    show linear and circular polarization respectively, while the
    solid line is the total intensity.  The mirror-like symmetry of
    the profile (see inset) implies that the pulses come from two
    crossings of a single cone of emission.  (Bottom) A rotating
    vector model (RVM) fit to the data (see \S\ref{sec:polar}).  See
    \citet{drb+04} for more details.
    \label{fig:polarization}}
\end{figure}

Fig.~\ref{fig:polarization} shows the 820\,MHz GBPP observations of
\A\ and the mirror-like symmetry of the pulse profile.  We have chosen
longitude $\phi=0$ as the center of this symmetry, which is visible in
both the total power and linearly polarized profiles.  Also shown is
the position angle versus pulse longitude for points having
$>$4$\sigma$ significance.  The longitude region
$-100^\circ<\phi<-70^\circ$ appears to contain an orthogonal emission
mode jump. The region $+90^\circ<\phi<+105^\circ$ may also be affected
by quasi-orthogonal modes.  We find the rotation measure of this
pulsar to be $-112.3\pm1.5$\,rad\,m$^{-2}$.

The mirror symmetry of the profile suggests that the emission is
coming from two traversals of a wide, hollow emission cone, rather
than from two opposite magnetic poles.  We have performed a rotating
vector model \citep[RVM;][]{rc69a} fit to the data which seems to
support this interpretation.  The fit gives values for the angle
between the magnetic and rotation axes ($\alpha$) and the closest
angle between the magnetic axis and the line of sight ($\beta$).  We
can also calculate the cone opening half-angle $\rho$, using the
longitude separation of the two peaks.  The fit produces two
solutions, one with $\alpha = 4^\circ \pm 3^\circ$ and one with
$\alpha = 90^\circ \pm 10^\circ$.  The large-$\alpha$ solution
requires a extremely large opening angle, $\rho \sim 90^\circ$.  We
favor the small-$\alpha$ solution, as it is more consistent with the
mirror symmetry, and permits more modest values of $\rho$.  In this
solution, $\beta$ is unconstrained, although we argue that it is
likely that $\beta \sim \rho$.

The polarimetric properties of \B\ are also interesting, though less
easily understood.  In addition to showing a remarkable intensity
variation throughout its orbit, this pulsar shows variation in its
degree of linear polarization.  The brightest ``window'' (I), just
before conjunction, shows a peak linear polarization fraction of
$\sim$14\%.  However, in the next bright window (II), just after
conjunction, the pulse is almost completely unpolarized ($\la$1\%).
From window I, we determine the rotation measure to be
$-118\pm12$\,rad\,m$^{-2}$, consistent with our value for \A.  More
details on these observations, the model fits, and their implications
are given in \citet{drb+04}.

\section{Eclipses of \A}
\label{sec:eclipses}

\begin{figure}
  \plotfiddle{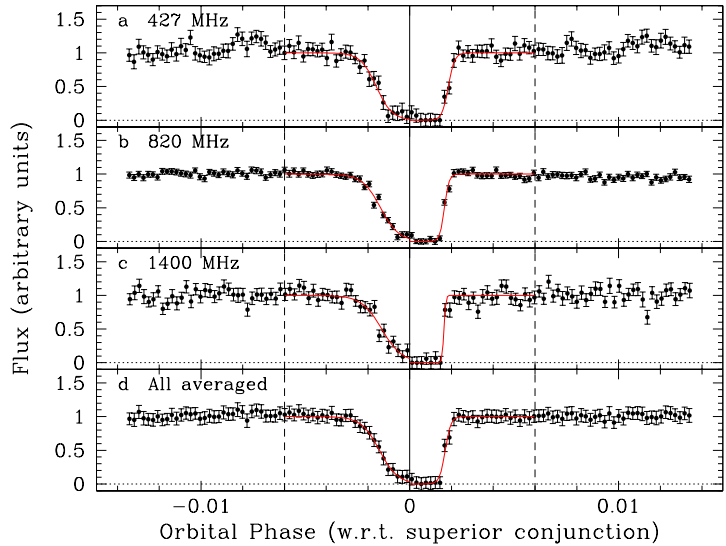}{7.5cm}{270}{38}{38}{-150}{230}
  \caption{
    Plots of the eclipses of \A\ at 427-, 820- and 1400-MHz with a
    time resolution of $\sim$2\,s.  The asymmetric ingress and egress
    of the eclipses and its near frequency independence are obvious.
    The deepest (i.e. most complete) part of the eclipse occurs just
    after conjunction.
    \label{fig:eclipses}}
\end{figure}

We used the BCPM data to study the eclipse of \A.  We folded data
obtained within 2\,min of \A's superior conjunction in 2-s intervals
using 256 phase bins and then cross-correlated each profile with a
template. The cross-correlation, done in the Fourier domain, assumes
the data profile $p$ has the form $p(j) = a + b \times s(j - \tau) +
n(j),$ for $1<j<256$, and where $s(j)$ is the template, $b$ is a scale
factor, $n(j)$ is the noise background, and $\tau$ is a temporal
offset.  Values of $a$, $b$ and $\tau$ were adjusted iteratively to
minimize a $\chi^2$ fit to each profile \citep{tay90a}.  We used the
scale factor $b$ and its standard deviation as a relative measure of
the pulsar's flux and its uncertainty, respectively.  Our resulting
eclipse light curves at three radio frequencies are presented in
Fig.~\ref{fig:eclipses}.  The eclipse properties are remarkably
independent of frequency, as previously noted qualitatively by
\citet{lbk+04}.  Because of its largely achromatic properties, we show
in Fig.~\ref{fig:eclipses}d an eclipse light curve obtained by
averaging those in panels a, b, and c.  A striking eclipse property is
that it is asymmetric at all frequencies in a similar way.  The
eclipse ingress is clearly longer in duration than the egress.

To quantify some of these properties, we fit the light curves with two
functions of the form $F(\phi) = (e^{(\phi - \phi_0)/w} + 1)^{-1}$,
where $F(\phi)$ is the flux at orbital phase $\phi$, defined such that
superior conjunction is at $\phi = 0$.  The best fits are shown as
solid curves in Fig.~\ref{fig:eclipses} and all yield values of
reduced-$\chi^2$ near unity.  Using the fit parameters, we
quantitatively show that the eclipse duration is independent of
frequency with a limit on a power-law index of $\sim$0.5; that the
ingress takes $\sim$3 times longer than egress, and that the eclipse
very likely lasts longer post-conjunction.  There is a hint that the
eclipse asymmetry grows with increasing radio frequency, but this is
not statistically significant.

A possible model for the \A\ eclipse is presented in these proceedings
by Arons et al. (see also Arons et al. in prep).  \A's wind likely
confines \B's magnetosphere on the side facing \A, causing it to
resemble a time-dependent variant of the Earth's magnetosphere.
Synchrotron absorption in a shock heated ``magnetosheath'' surrounding
and containing \B's magnetosphere is then responsible for the eclipse.
Additional details on the data analysis and the eclipse model can be
found in \citet{krb+04}.

\section{Scintillation Velocity}
\label{sec:scint}

\begin{figure}
  \plotfiddle{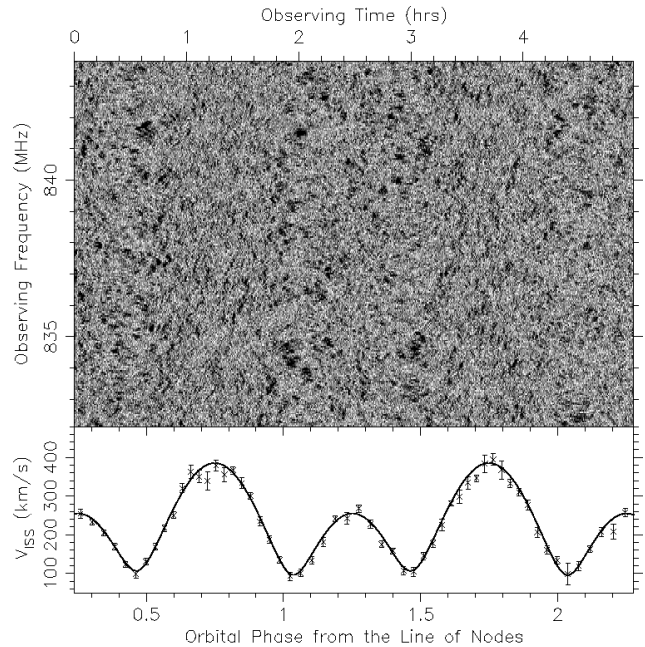}{8.0cm}{0}{55}{55}{-180}{-100}
  \caption{
    Dynamic spectra (top) and a scintillation velocity model fit
    (bottom) for the 820-MHz SPIGOT observation.  Only the top 1/4 of
    the band is plotted.  The fit implies that 0737 is moving across
    the sky at $\sim$140\,km\,s$^{-1}$.  See \S\ref{sec:scint} and
    \citet{rkr+04} for more details.
    \label{fig:scint}}
\end{figure}

The high SNR of the GBT data at 1400\,MHz and the fantastic frequency
resolution of the SPIGOT at 820\,MHz allowed us to make
``self-calibrating'' measurements of the orbital variation of the
diffractive scintillation timescale $\Delta t_d$ for \A\ and hence,
the so-called scintillation velocity $V_{\rm ISS} \propto 1/\Delta
t_d$ \citep[e.g.][]{cr98}.  Closely following the methodology of
\citet*{obvs02}, we computed dynamic spectra for \A\ for the two
observations (see Fig.~\ref{fig:scint}), and then using the known
(from timing) orbital motion of \A, fit a model consisting of four
parameters: the orbital inclination $i$, the systemic velocity
parallel- ($V_{\rm plane}$) and perpendicular-to ($V_{\rm perp}$) the
line of nodes on the sky, and the scaling parameter $\kappa$.  Our
results indicate that 0737 is moving across the sky at $V_{\rm
  ISS}$$\simeq$140.9$\pm$6.2\,km\,s$^{-1}$, with component veocities
of $V_{\rm plane}$$\simeq$96.0$\pm$3.7\,km\,s$^{-1}$ and $V_{\rm
  perp}$$\simeq$103.1$\pm$7.7\,km\,s$^{-1}$.  When combined with the
timing fits described in \S\ref{sec:timing}, the best-fit inclination
is $i=88\fdg7\pm0\fdg9$.

Using a simple kinematic argument and estimates of the characteristics
of 0737 before the second supernova \citep{dv04,wk04}, the large
$V_{\rm perp}$ implies that \B\ was born with a kick speed of
$\ga$100\,km\,s$^{-1}$.  Future measurements of the scintillation
velocity should allow us to remove the contaminating effects of the
Earth's motion around the Sun and the differential rotation of the
Galaxy.  A precise systemic velocity in combination with a
VLBA-determined proper motion will provide an accurate geometric
distance to the system (which could be crucial for high precision
tests of gravitational theories; see the contribution by Kramer in
this volume). See \citet{rkr+04} for more details.

\section{Pulse Fluctuations and Morphology}
\label{sec:pulses}

\begin{figure}
  \plotfiddle{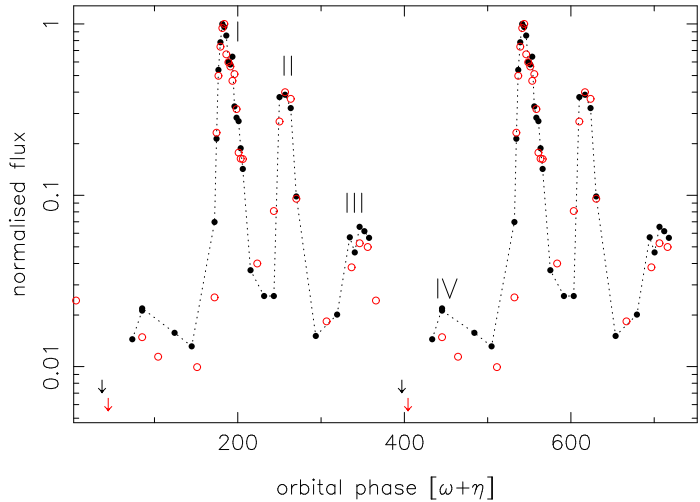}{6.0cm}{270}{40}{40}{-150}{225}
  \caption{
    820- (filled circles) and 1400-MHz (open circles) orbital
    lightcurve of \B.  See \S\ref{sec:pulses} and \citet{rdr+04} for
    details.
    \label{fig:lightcurve}}
\end{figure}

Due to the superior sensitivity of the GBT, we detected \B\ at good
signal-to-noise levels almost throughout the orbit.
Fig.~\ref{fig:lightcurve} shows the normalized flux received from \B\ 
as a function of the true anomaly (referenced from the line of nodes)
$\phi_{\rm orb}$, at 820- (filled circles) and 1400-MHz (open
circles). The light curve shows significant systematic variation, as
first shown by \citet{lbk+04}, and four prominent phase windows where
the pulsar is bright (indicated by I, II, III \& IV; see also
Fig.~\ref{fig:Btiming}).  In the range of $\phi_{\rm orb} \sim 6\deg$
to $65\deg$, the pulsar appears to exhibit what may be interpreted as
an eclipse, where we did not detect any signal within our sensitivity
limits. In fact, around the eclipse boundary, the reduction in SNR was
so significant that the pulsar became undetectable within a few
rotation periods of \B.  The Arons et al.~magnetopause model (these
proceedings; 2004, in prep) may be able to explain at least some of
\B's systematic flux variation (including the eclipse), as well as
\A's eclipse.  \citet[][see also the contribution from Jenet in this
volume]{jr04} propose an alternate model for \B's flux variation where
\B\ brightens when it is illuminated by the cone of (presumably
high-energy or particle) emission from \A.

Both \A\ and \B\ exhibit significant intrinsic pulse to pulse
variations. The measured modulation index $m$ at 820\,MHz as a
function of pulse longitude for \A\ is in the range 0.6$-$2.0.  \B's
profile structure is much more complex, where even the average profile
changes systematically as a function of $\phi_{\rm orb}$ \citep[as
first discussed by][]{lbk+04}, presumably indicating a change in the
emission beam morphology.  In windows I and II, values of $m$ are in
the range 1$-$1.5 and 1$-$2, respectively. These values are
significantly greater than what is expected from interstellar
scattering ($\sim$7\%), given the measured frequency and temporal
dimensions of scintles and their filling factor \citep{rkr+04}.  The
spectral shape of these intrinsic fluctuations are ``noiselike'' (i.e.
white), indicating the lack of any organized drifting patterns.  We
present a much more detailed study of the above mentioned properties
in \citet{rdr+04}.

\section{Conclusion}

The fantastic and highly-varied science to come from these five
observations gives some indication of how important a role the GBT
will play in future work on 0737.  These observations provide the best
constraints yet available on the inclination of the orbits, the mass
ratio of the pulsars, the geometry of \A's radio emission, the
asymmetries and achromaticity of \A's eclipses (and the nature of \B's
magnetosphere that cause them), the orbital modulation of \B's pulsed
flux, and the systemic velocity of 0737 and the kick that caused it.

The high gain of the telescope, the availability of many sensitive and
useful receivers, and the new pulsar backends (including the SPIGOT
and two coherent de-dispersion systems) will insure that these are
only the first of a long series of GBT-based results from the
double-pulsar.


\begin{thebibliography}{13}
\expandafter\ifx\csname natexlab\endcsname\relax\def\natexlab#1{#1}\fi

\bibitem[{{Burgay} {et~al.}(2003){Burgay}, {D'Amico}, {Possenti}, {Manchester},
  {Lyne}, {Joshi}, {McLaughlin}, {Kramer}, {Sarkissian}, {Camilo}, {Kalogera},
  {Kim}, \& {Lorimer}}]{bdp+03}
{Burgay}, M., {D'Amico}, N., {Possenti}, A., {Manchester}, R.~N., {Lyne},
  A.~G., {Joshi}, B.~C., {McLaughlin}, M.~A., {Kramer}, M., {Sarkissian},
  J.~M., {Camilo}, F., {Kalogera}, V., {Kim}, C., \& {Lorimer}, D.~R. 2003,
  Nature, 426, 531

\bibitem[{{Cordes} \& {Rickett}(1998)}]{cr98}
{Cordes}, J.~M. \& {Rickett}, B.~J. 1998, \apj, 507, 846

\bibitem[{{Demorest} {et~al.}(2004){Demorest}, {Ramachandran}, {Backer},
  {Ransom}, P.~{Kaspi}, {Arons}, \& {Spitkovsky}}]{drb+04}
{Demorest}, P., {Ramachandran}, R., {Backer}, D.~C., {Ransom}, S.~M.,
  P.~{Kaspi}, V.~M., {Arons}, J., \& {Spitkovsky}, A. 2004, \apj, submitted
  (astro-ph/0402025)

\bibitem[{{Dewi} \& {van den Heuvel}(2004)}]{dv04}
{Dewi}, J.~D.~M. \& {van den Heuvel}, E.~P.~J. 2004, \mnras, 349, 169

\bibitem[{{Jenet} \& {Ransom}(2004)}]{jr04}
{Jenet}, F.~A. \& {Ransom}, S.~M. 2004, Nature, in press

\bibitem[{{Kaspi} {et~al.}(2004){Kaspi}, {Ransom}, {Backer}, {Ramachandran},
  {Demorest}, {Arons}, \& Spitkovsky}]{krb+04}
{Kaspi}, V.~M., {Ransom}, S.~M., {Backer}, D.~C., {Ramachandran}, R.,
  {Demorest}, P., {Arons}, J., \& Spitkovsky, A. 2004, \apj, submitted
  (astro-ph/0401614)

\bibitem[{{Lyne} {et~al.}(2004){Lyne}, {Burgay}, {Kramer}, {Possenti},
  {Manchester}, {Camilo}, {McLaughlin}, {Lorimer}, {D'Amico}, {Joshi},
  {Reynolds}, \& {Freire}}]{lbk+04}
{Lyne}, A.~G., {Burgay}, M., {Kramer}, M., {Possenti}, A., {Manchester}, R.~N.,
  {Camilo}, F., {McLaughlin}, M.~A., {Lorimer}, D.~R., {D'Amico}, N., {Joshi},
  B.~C., {Reynolds}, J., \& {Freire}, P.~C.~C. 2004, Science, 303, 1153

\bibitem[{{Ord} {et~al.}(2002){Ord}, {Bailes}, \& {van Straten}}]{obvs02}
{Ord}, S.~M., {Bailes}, M., \& {van Straten}, W. 2002, \apjl, 574, L75

\bibitem[{Radhakrishnan \& Cooke(1969)}]{rc69a}
Radhakrishnan, V. \& Cooke, D.~J. 1969, Ann. Phys. (Leipzig), 3, 225

\bibitem[{{Ramachandran} {et~al.}(2004){Ramachandran}, {Backer}, {Demorest},
  {Ransom}, \& {Kaspi}}]{rdr+04}
{Ramachandran}, R., {Backer}, D.~C., {Demorest}, P., {Ransom}, S.~M., \&
  {Kaspi}, V.~M. 2004, \apj, submitted

\bibitem[{{Ransom} {et~al.}(2004){Ransom}, {Kaspi}, {Ramachandran}, {Demorest},
  {Backer}, {Pfahl}, {Ghigo}, \& {Kaplan}}]{rkr+04}
{Ransom}, S.~M., {Kaspi}, V.~M., {Ramachandran}, R., {Demorest}, P., {Backer},
  D.~C., {Pfahl}, E.~D., {Ghigo}, F.~D., \& {Kaplan}, D.~L. 2004, \apj,
  submitted (astro-ph/0404149)

\bibitem[{Taylor(1990)}]{tay90a}
Taylor, J.~H. 1990, in Impact of Pulsar Timing on Relativity and Cosmology, ed.
  D.~C. Backer (Berkeley: Center for Particle Astrophysics), m1

\bibitem[{{Willems} \& {Kalogera}(2004)}]{wk04}
{Willems}, B. \& {Kalogera}, V. 2004, \apjl, 603, L101

\end{thebibliography}

\end{document}